\documentclass[preprint,amsmath,amssymb,prl, letter]{revtex4-2}
\usepackage{graphicx}
\usepackage{setspace,graphicx}%, subfig}
\usepackage{amsmath,amssymb,amsfonts,amsthm,braket,verbatim,hyperref}

\usepackage{mathrsfs}
\usepackage{xcolor}
\usepackage{epstopdf}
\usepackage{tikz,pgfplots}
\usepgfplotslibrary{fillbetween}

\usepackage{parskip}
\usetikzlibrary{matrix}
\pgfplotsset{compat=1.12}
\usepgfplotslibrary{fillbetween}
\usetikzlibrary{patterns}
\usepgfplotslibrary{external}
\let\vec\mathbf

\begin{document}

%\title{Emergent canonical variables and a generalized uncertainty relation between quantum fluctuations constrain the energy of many-body systems}
%\title{A generalized uncertainty constraint on quantum fluctuations of interacting many-particle systems -- energy functionals and coherence of ground states}
%\title{Charge density constrains quantum fluctuations of effective canonical variables in interacting, many-body quantum systems} %  determining interacting many-particle systems through local constraints on quantum fluctuations}
%\title{Charge density enables the Hohenberg-Kohn theorem by constraining quantum fluctuations of canonical variables in interacting, many-body quantum systems} %  determining 
\title{Uncertainty relations for the Hohenberg-Kohn theorem}
\author{Purnima Ghale}
\email{pg472@cornell.edu}
 \affiliation{Cornell High Energy Synchrotron Source \\ Cornell University, Wilson Lab, Synchrotron Drive, Ithaca, NY, 14850}

\date{\today}% It is always \today, today,

\begin{abstract}
How does charge density constrain many-body wavefunctions in nature? The Hohenberg-Kohn theorem for non-relativistic, interacting many-body Schr\"odinger systems is well-known and was proved using \emph{reductio-ad-absurdum}; however, the physical mechanism or principle which enables this theorem in nature has not been understood. Here, we obtain effective canonical operators in the interacting many-body problem -- (i) the local electric field, which mediates interaction between particles, and contributes to the potential energy; and (ii) the particle momenta, which contribute to the kinetic energy. The commutation of these operators results in the charge density distribution. Thus, quantum fluctuations of interacting many-particle systems are constrained by charge density, providing a mechanism by which an external potential, by coupling to the charge density, tunes the quantum-mechanical many-body wavefunction. As an initial test, we obtain the functional form for total energy of interacting many-particle systems, and in the uniform density limit, find promising agreement with Quantum Monte Carlo simulations. 
\end{abstract}

\maketitle
\newpage
%Despite the universal nature of the dependence between charge density and ground states, however, the underlying mechanistic structure through which charge density determines many-electron systems is not fully understood.
The Hohenberg-Kohn theorem states that for non-relativistic many-electron systems described by the Schr\"odinger equation, once the external potential and charge density are fixed, the system is fully determined\cite{hohenberg1964inhomogeneous}. Specifically, we consider interacting many-body Schr\"odinger Hamiltonians in atomic units: 
\begin{subequations}
\begin{equation}
\hat{\mathscr{H}} = \underbrace{\sum_\alpha \frac{\hat{\vec{p}}_\alpha^2}{2}}_{\hat{T}} + \underbrace{\frac{1}{2} \sum_{\alpha \neq \beta} \frac{1}{|\hat{\vec{x}}_\alpha - \hat{\vec{x}}_\beta|}}_{\hat{V}_{ee}} + V_{ext}(\vec{r})
\label{eq:hamiltonian}
\end{equation}
Let $\{\hat{\vec{x}}_{\alpha=1}, \ldots, \hat{\vec{x}}_N\}$, $\{ \hat{\vec{p}}_{\alpha=1}, \ldots, \hat{\vec{p}}_N\}$, and $\{ \vec{r}\}$ denote particle coordinates, particle momenta, and spatial coordinate on which local external potential $V_{ext}(\vec{r})$ is defined, respectively; $\hat{T}$ and $\hat{V}_{ee}$ denote kinetic and interaction energy operators, and many-body wavefunctions are denoted by $\Psi(\vec{x}_1, \ldots, \vec{x}_N)$. The universal density functional, $F[n(\vec{r})]$, maps the complicated many-body wavefunction to a scalar field in three dimensions, and is defined by 
\begin{equation}
F[n(\vec{r})]= \text{min}_{\Psi \rightarrow n(\vec{r})}  \left \{ \braket{\Psi | \hat{T} + \hat{V}_{ee}|\Psi}  \right \}
\end{equation} 
\end{subequations}
Mathematically, separate bounds on the kinetic energy, $\braket{\Psi | \hat{T}| \Psi}$, and the interaction energy, $\braket{\Psi | \hat{V}_{ee}| \Psi}$, have been obtained as functionals of charge density \cite{chan1999optimized,lieb1979lower,lieb1981improved,liebkineticprl}. In addition, the Kohn-Sham ansatz\cite{sham1966one,kohn1965self} can be used to understand weakly interacting systems with approximate exchange-correlation functionals, and a hierarchy of functionals of increasing complexity can be constructed\cite{perdew2001jacob,mardirossian2015mapping,tao2003climbing}. But there are some limitations to prior work. One limitation is that constraints on the sum of operators, $\braket{\Psi | \hat{T} + \hat{V}_{ee} | \Psi}$, remain elusive. Another, is that not all particle densities correspond to quantum many-body ground states \cite{kohn1983v,yang2004potential,lieb1983eh}, and we cannot fully understand which densities are $V$-representable without direct mechanistic insight into $F[n(\vec{r})]$. Furthermore, while the Kohn-Sham ansatz\cite{kohn1965self} is highly effective, the non-factorizability of many-body wavefunctions \cite{nonfermiliquid} and ambiguity on the accuracy of exchange-correlation functionals\cite{straying, dilemma_on, comment_on}, pose fundamental problems for quantum many-body systems with strong interactions. The benchmark calculations for many strongly-correlated systems are based on Quantum Monte Carlo (QMC)\cite{ceperley1978ground,ceperley1980ground}, but even then, the number of possible fermion nodal surfaces or antisymmetric shapes of many-electron wavefunctions remains a challenge\cite{mitas2006structure,ceperley1991fermion,carlson2011auxiliary,sunko2016natural}. 

Despite difficulties in understanding interacting many-particle systems, however, the Hohenberg-Kohn theorem is simple and universally applicable, shown via \textit{reductio-ad-absurdum} in \cite{hohenberg1964inhomogeneous}. Our goal, therefore, is to understand the physical reason behind the success of this theorem -- in particular, mechanistically, why/how does nature implement the Hohenberg-Kohn theorem?  

\begin{figure}[htb!]
\centering
{\includegraphics[width=0.9\linewidth]{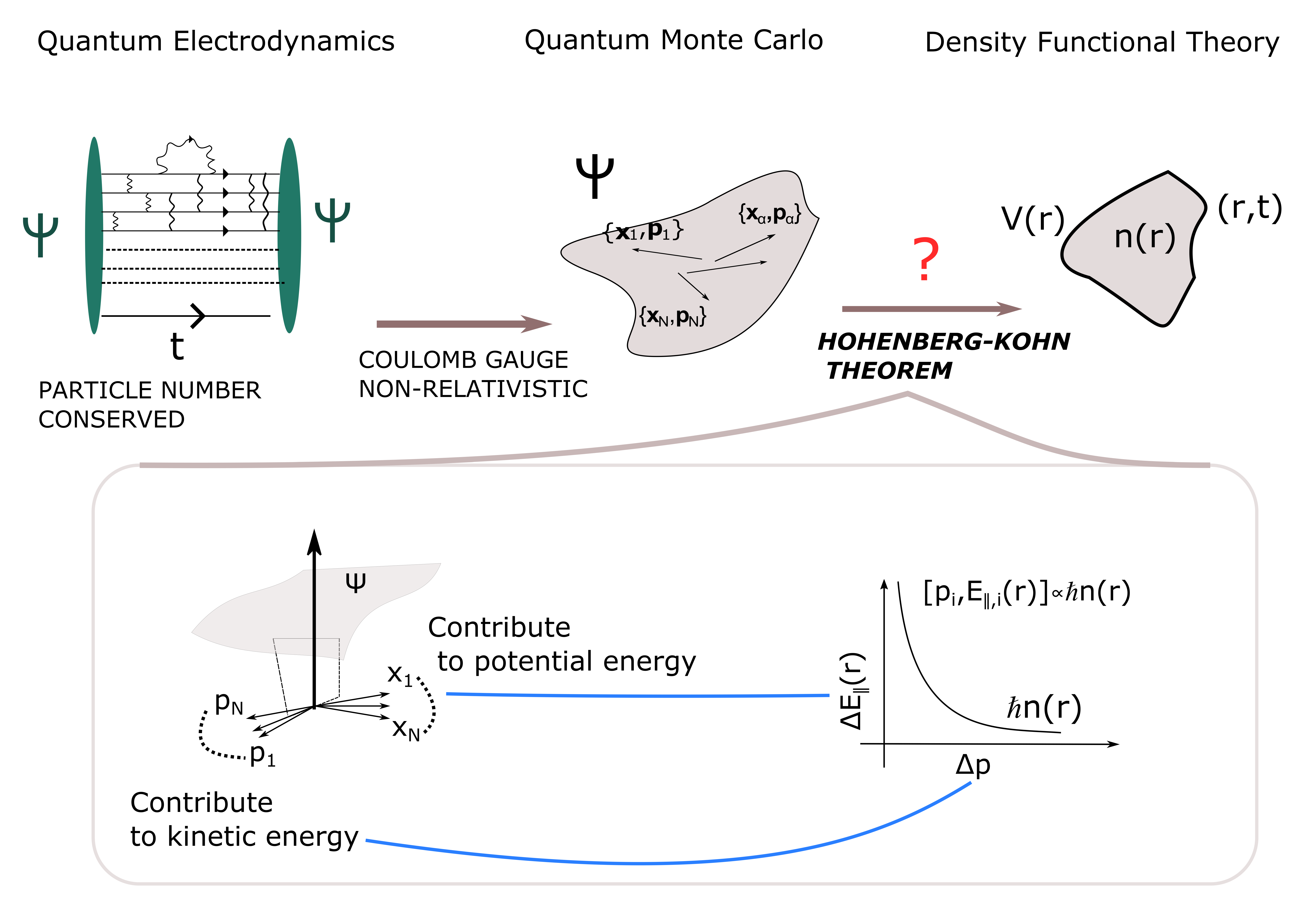}} 
\caption{\label{fig:fig1} In three conceptual leaps, relativistic particles dressed by photons can be investigated using a theory determined entirely in terms of charge density. Recall that longitudinal photons mediate interactions in Quantum Electrodynamics, and choosing the radiation gauge allows us to obtain the Quantum Monte Carlo Hamiltonian of Equation~\ref{eq:hamiltonian}. The Hohenberg-Kohn theorem then maps the many-body wavefunction $\Psi$ to a scalar field $n(\vec{r})$, but the physical mechanism that enables this mapping was unknown. We propose the existence of effective canonical operators in interacting many-particle systems: particle momenta and local electric fields, with their commutation resulting in the local charge density. This operator algebra and the implied uncertainty relation constrain quantum fluctuations of interacting many-particle systems.}
\end{figure}

Figure~\ref{fig:fig1} summarizes our main result. Longitudinal photons in radiation gauge of quantum electrodynamics correspond to the Coulomb interaction, $V_{ee} \sim \frac{1}{|\vec{x}_\alpha - \vec{x}_\beta|}$, of Equation~\ref*{eq:hamiltonian}. $\braket{\hat{V}_{ee}}$ can therefore be replaced by the integrated energy density of the electric field. Moreover, we find that the local electric field and particle momentum are relevant canonical operators of the system: (i) these operators contribute to the potential and kinetic energy respectively, (ii) their quantum fluctuations contribute to the quantum-mechanical corrections to total energy, and (iii) commutation of the two operators results in the charge density, such that charge density constrains quantum fluctuations of the many-body system. We will first derive the relationship between operators shown in Figure~\ref{fig:fig1}, and use that relation to obtain a functional form for total energy of interacting many-particle systems. For the jellium model, the functional form obtained in this way agrees with reference Quantum Monte Carlo (QMC) simulations. %For uniformly dense systems, the functional form for total energy is parameterized against four types of quantum many-body states investigated using Quantum Monte Carlo (QMC) in \cite{ceperley1980ground}, and we find promising agreement despite differences in particle statistics, spin polarization or crystallinity. %The parameterizations succeed in representing the energies and transitions from \cite{ceperley1980ground}, except the paramagnetic to ferromagnetic fermionic transition. This happens for the right reason -- spin-orbit coupling is required to understand the split but is not present in the Hamiltonian used (\textit{cf.} Equation~\ref{eq:hamiltonian}). %Note, however, that minimum uncertainty states are coherent states, distinct from eigenstates of the Hamiltonian in Equation~\ref{eq:hamiltonian}, although ground states are often both coherent and eigenstates. Recalling that faster degrees of freedom in the ambient electromagnetic field are not included in Equation~\ref{eq:hamiltonian}, our results suggest that the non-relativistic many-body quantum states we observe and engineer by applying local external potentials, are likely to be coherent many-body states that survive in the presence of coupling to the electromagnetic bath. 
%\section{Derivation of constraint on fluctuations}

\textbf{\emph{Interaction energy in terms of longitudinal electric fields}} -- We first show that the electron-electron interaction is the square of the energy density of the electric field: 
\begin{equation}
\braket{\Psi | \hat{V}_{ee} | \Psi} = \frac{1}{8\pi} \int_{\vec{r}} \braket{\Psi | \hat{\vec{E}}_{\parallel}(\vec{r}) \cdot \hat{\vec{E}}_{\parallel}(\vec{r})|\Psi}
%\hat{H} = -\sum_{\alpha}\frac{\hat{p}_{\alpha}^2}{2} + \int_r E_{\parallel}(\vec{r}) \cdot E_{\parallel}(\vec{r}) + \underbrace{\int_r V_{ext}(\vec{r}) n(\vec{r})}_{\text{fixed}}
\end{equation}
where $\hat{\vec{E}}_{\parallel}(\vec{r})$ denotes the electric field generated by particles with coordinates $\{ \vec{x}_1, \ldots, \vec{x}_\alpha, \ldots, \vec{x}_N\} $ at point $\vec{r}$, and is given by the gradient of the local potential $\hat{\vec{E}}_{\parallel,j}(\vec{r}) = -\partial_{r_j} \sum_\beta \frac{1}{| \vec{r} - \vec{x}_{\beta}|}$. Let us now consider:
\begin{subequations}
\begin{equation}
 \int_r\braket{\Psi | \hat{\vec{E}}_{\parallel}(\vec{r}) \cdot \hat{\vec{E}}_{\parallel} (\vec{r}) | \Psi} = \int_r  \braket{\Psi | \sum_{\alpha,\beta} \partial_{r_j} \frac{1}{|\vec{r} - \vec{x}_\alpha|} \cdot \partial_{r_j} \frac{1}{|\vec{r} - \vec{x}_\beta|} | \Psi} \end{equation}
Integrating by parts in $\vec{r}$, and setting surface terms to zero, we obtain 
\begin{equation} 
= -\sum_{\alpha,\beta} \int_r \braket{\Psi |  \frac{1}{|\vec{r} - \vec{x}_\alpha|} \nabla^2_r \frac{1}{|\vec{r} - \vec{x}_\beta|} | \Psi} 
\end{equation}
Using $\nabla_r^2\frac{1}{|\vec{r} - \vec{x}_\beta|} = -4\pi \delta^{(3)} (\vec{r} - \vec{x}_\beta)$, results in
\begin{equation}
=\sum_{\alpha,\beta} \int_r \braket{\Psi |  \frac{4 \pi}{|\vec{r} - \vec{x}_\alpha|} \delta^{(3)}(\vec{r} - \vec{x}_\beta) | \Psi} = 4\pi \sum_{\alpha,\beta} \braket{\Psi |  \frac{1}{|\vec{x}_\beta - \vec{x}_\alpha|} | \Psi} 
\end{equation}
\begin{equation}
 = 4 \pi \sum_{\alpha \neq \beta} \braket{\Psi |  \frac{1}{|\vec{x}_\beta - \vec{x}_\alpha|} | \Psi} + 4\pi N \underbrace{ \braket{\Psi |\int_{r}   \frac{\delta(\vec{r} - \vec{x}_{1})}{ |\vec{r} - \vec{x}_1|} | \Psi}}_{\Lambda}
\end{equation}
The last term, denoting self-interaction, simplifies to a renormalizable constant, ${\left( \sqrt{2\pi}\right)^3} \int_{\vec{k}}  \frac{1}{|\vec{k}|^2} \rightarrow \Lambda$ and depends on the UV-cutoff used for regularization, alternatively, also see \cite{note2b}. We have thus obtained the interaction energy in terms of longitudinal electric fields, as follows.
\begin{equation}
 \int_r\braket{\Psi | \hat{\vec{E}}_{\parallel}(\vec{r}) \cdot \hat{\vec{E}}_{\parallel} (\vec{r}) | \Psi} = 8\pi \braket{\Psi | \hat{V}_{ee}| \Psi} + 4\pi N \Lambda \label{eq:E2_quant}
\end{equation}
\end{subequations}
 A similar analysis (integration by parts, setting surface terms to zero, and using $\nabla^2_{r} \frac{1}{|\vec{r} - \vec{x}|} = \delta^{(3)}(\vec{r} - \vec{x})$) results in the connection between the energy density of the classically defined electric field, and the electrostatic interaction between charges, as follows.
\begin{equation} \int_{\vec{r}} \braket{\Psi|\hat{\vec{E}}_{\parallel}(\vec{r})|\Psi} \cdot \braket{\Psi|\hat{\vec{E}}_{\parallel}(\vec{r})|\Psi} = 8\pi \int_{\vec{r}, \vec{r}'} \frac{1}{2}\frac{n(\vec{r}) n(\vec{r}')}{|\vec{r} - \vec{r}'|}
\label{eq:E2_class}
\end{equation}
Using Eqs~\ref{eq:E2_quant} and \ref{eq:E2_class}, the interaction energy, $\braket{\Psi | \hat{V}_{ee} | \Psi}$, separates into a classical, electrostatic contribution, and contributions due to quantum fluctuations of $\hat{\vec{E}}_{\parallel}(\vec{r})$. 
\begin{equation}
\braket{\hat{V}_{ee}}  = \frac{1}{8\pi} \braket{ \hat{\vec{E}}_{\parallel}(\vec{r})} \cdot \braket{\hat{\vec{E}}_{\parallel}(\vec{r}) } + \frac{1}{8\pi} \int_{\vec{r}}  \Delta \vec{E}^2_{\parallel}(\vec{r}) 
\end{equation}
\begin{equation}
\braket{\hat{V}_{ee}}  = \frac{1}{2} \int_{\vec{r}, \vec{r}'} \frac{n(\vec{r}) n(\vec{r}')}{|\vec{r} - \vec{r}'|} + \frac{1}{8\pi} \int_{\vec{r}} \Delta \vec{E}^2_{\parallel}(\vec{r}) 
\end{equation}
where $\Delta \vec{E}^2_{\parallel}(\vec{r}) = \braket{\hat{\vec{E}}_{\parallel}(\vec{r}) \cdot \hat{\vec{E}}_{\parallel}(\vec{r}) } -  \braket{\hat{\vec{E}}_{\parallel}(\vec{r}) } \cdot \braket{\hat{\vec{E}}_{\parallel}(\vec{r}) }$.%\int_{\vec{r}, \vec{r}'} \frac{1}{2}\frac{n(\vec{r}) n(\vec{r}')}{|\vec{r} - \vec{r}'|}  +

\textbf{\emph{Kinetic energy of many-particle quantum systems}} -- As for contributions to kinetic energy, let us assume for convenience that the total momentum, $\braket{\hat{\vec{p}}} = 0$, i.e. the many-particle system is not undergoing translation, or the reference frame is chosen to be so. Particle indistinguishability then implies\cite[see p. 320]{lieb1988Denmark} $ \braket{\hat{\vec{p}}_{1,i}} = \braket{\hat{\vec{p}}_{2,i}} =\ldots = \braket{\hat{\vec{p}}_{N,i}}=0$, and the kinetic energy is the sum of fluctuations of momentum, 
\begin{equation}
\braket{\hat{T}} = \sum_{\alpha=1}^N \frac{\Delta \vec{p}_{\alpha}^2}{2} 
\label{eq:ke}
\end{equation}
where $\Delta \vec{p}^2_\alpha = \braket{\hat{\vec{p}}_\alpha ^2} - \braket{\hat{\vec{p}}_\alpha}^2 $. The universal interacting many-body functional, $F[n(\vec{r})]$, in terms of classical electrostatic energy and quantum fluctuations is
\begin{equation}
F[n(\vec{r})] = \braket{\hat{T} + \hat{V}_{ee}} = \sum_{\alpha} \frac{\Delta  \vec{p}_{\alpha}^2}{2} + \frac{1}{2}\int_{r,r'} \frac{n(\vec{r}) n(\vec{r'})}{|\vec{r} - \vec{r'}|}  + \frac{1}{8\pi}\int_r \Delta E_{\parallel}^2(\vec{r})
\end{equation}

\textbf{\emph{Effective commutation relation}} -- We have thus mapped the many-body Schr\"odinger hamiltonian to classical and quantum contributions, such that the quantum contributions are given by 
\begin{subequations}
\begin{equation}
 \mathscr{E}_{quantum}= \sum_{\alpha}\frac{\Delta \vec{p}_{\alpha}^2}{2} + \frac{1}{8\pi}\int_{\vec{r}}\Delta \vec{E}^2_\parallel(\vec{r})
\end{equation}
Inspecting the relationship between the momentum and electric field operators via commutation, we obtain
\begin{equation} [\hat{\vec{p}}_{j,\alpha}, \hat{\vec{E}}_{\parallel,k}(\vec{r})] \Psi = \frac{\hbar}{i} \left(\partial_{x_{j,\alpha}} \vec{E}_{\parallel,k}(\vec{r}) \right)  \Psi = -\frac{\hbar}{i} \left( \partial_{x_{j,\alpha}} \partial_{r_k} \sum_\beta \frac{1}{|\vec{r} - \vec{x}_\beta|}\right)\Psi 
\end{equation}
Changing the order of differentiation, $\partial_{x_{j,\alpha}}$ selects only $\alpha = \beta$ from summation over $\beta$; furthermore, $\partial _{x_{j}} \frac{1}{|\vec{r} - \vec{x}|} = \partial _{r_{j}} \frac{1}{|\vec{r} - \vec{x}|}$. Thus, we obtain:
\begin{equation}
[\hat{\vec{p}}_{j,\alpha}, \hat{\vec{E}}_{\parallel,k}(\vec{r})] \Psi = \frac{\hbar}{i} \left( -\partial_{r_j} \partial_{r_k} \frac{1}{|\vec{r} - \vec{x}_\alpha|} \right) \Psi 
\label{eq:commutation3} \end{equation}  
Using $j=k$ terms, summation over repeated indices, and $-\nabla_r^2 \frac{1}{|\vec{r} - \vec{x}|} = 4\pi \delta^{(3)}(\vec{r} - \vec{x})$, we obtain:
\begin{equation}
[\hat{\vec{p}}_{j,\alpha}, \hat{\vec{E}}_{\parallel,j}(\vec{r})] \Psi = \frac{\hbar}{i} \left( -\nabla^2_r \frac{1}{|\vec{r} - \vec{x}_\alpha|} \right) \Psi   = \frac{4\pi \hbar}{i} \delta^{(3)}(\vec{r} - \vec{x}_\alpha)\end{equation}  
Summation over particle labels, $\alpha$ gives:
\begin{equation}
\sum_{\alpha} [\hat{\vec{p}}_{\alpha, j}, \hat{\vec{E}}_{\parallel, j} (\vec{r})]\Psi = \frac{4\pi \hbar}{i} \hat{n}(\vec{r})
\end{equation}
From particle indistinguishability, we obtain:
\begin{equation}
N [\hat{\vec{p}}_{j}, \hat{\vec{E}}_{\parallel, j} (\vec{r})]\Psi = \frac{4\pi \hbar}{i} \hat{n}(\vec{r})
\end{equation}
\end{subequations}
\begin{subequations}
\textbf{\emph{Constraint on quantum fluctuations of canonical variables}} --  Let us now constrain quantum fluctuations. Applying the Cauchy-Schwartz inequality following Robertson\cite{robertson1929uncertainty}, the commutation relations in Equation~\ref{eq:commutation3} imply the following relation between quantum fluctuations:
 \begin{equation}
 \Delta \vec{p}_{j,\alpha} \Delta \vec{E}_{\parallel,k} \ge \frac{\hbar}{2} \left| \braket{\Psi | -\partial_{r_j} \partial_{r_k} \frac{1}{|\vec{r} - \vec{x}_\alpha|} | \Psi}\right|
 \end{equation}
%\begin{equation}[\hat{p}_{j,\alpha}, \hat{E}_j(\vec{r})] \Psi =  \end{equation} 
%, results in a relationship between fluctuations of the kind, \begin{equation} \Delta {\vec{p}_{j,\alpha}} \Delta {\vec{E}_{\parallel,j}(\vec{r})}  \ge \frac{\hbar}{2} \left|\braket{\Psi|  -\partial^2_{\vec{r}_{j}} \frac{1}{|\vec{r} - \vec{x}_\alpha|}   | \Psi} \right| \end{equation} 
Using $|g(x)| \ge g(x)$, summing over three dimensions, and using the Poisson relation between potential and charge density, again, leads to  
\begin{equation} \Delta {\vec{p}_{j,\alpha}} \Delta {\vec{E}_{\parallel,j}(\vec{r})}  \ge 2\pi \hbar \braket{\hat{n}_{\alpha}(\vec{r})} \end{equation} Summation over particle labels, assuming isotropy of fluctuations, \begin{equation} \Delta \vec{p}^2_{x} = \Delta \vec{p}^2_{y} = \Delta \vec{p}^2_{z} = \Delta p^2 \end{equation}
\begin{equation} \Delta \vec{E}^2_{\parallel,x}(\vec{r}) = \Delta \vec{E}^2_{\parallel,y}(\vec{r}) = \Delta \vec{E}^2_{\parallel,z}(\vec{r}) = \Delta {E}_{\parallel}^2(\vec{r}) \end{equation}
and using particle indistinguishability, we get  
\begin{equation}
3N \Delta {{p}} \Delta{{E}_\parallel(\vec{r})} \ge  2\pi \hbar \, n(\vec{r}) \label{eq:uncertain}
\end{equation}
{Equation~\ref{eq:uncertain} suggests that by fixing $n(\vec{r})$, we are essentially realizing a generalized uncertainty constraint on quantum fluctuations of the many-particle system. Note that we are constraining fluctuations of longitudinal photons, which on their own have negative normalization in field theory \cite{ryder1996quantum}. It is therefore better to leave the inequality indeterminate for now as follows. 
\begin{equation} 
3N \Delta p \Delta E_{\parallel}(\vec{r}) \sim 2 \pi \hbar \, n(\vec{r}) 
\label{eq:uncertainb}
\end{equation} %To obtain positive normalization, an explicitly covariant treatment of the electromagnetic field, with compensating timelike photons (and transverse components) added to the Hamiltonian, would be necessary.
}
\end{subequations}

%\section{Energy of many-particle systems}
%So far, we have related fluctuations of momenta and electric fields to the charge density. Before we proceed further, recall that in the quantum harmonic oscillator, coherent states that satisfy canonical uncertainty relations are distinct from eigenstates of the Hamiltonian. Such coherent states evolve most classically, are robust to noise, and the ground state is both an eigenstate and a coherent state. Thus, the uncertainty relations we have derived suggest that for a given charge density, there exist a multiplicity of coherent states, with different symmetry under coordinate exchange, spin-polarization, crystallinity etc., that nevertheless satisfy the uncertainty relations. %sa constraint on fluctuations of operators, agnostic to symmetries in the wavefunction that are not modified by the hamiltonian. %of these coherent states, the ground state will have the lowest energy. 
\textbf{\emph{Effect of uncertainty relation on energy as a functional of $n(\vec{r})$}} -- \begin{subequations} Equation~\ref{eq:uncertainb} will now be used to obtain a density functional expression for total energies of many-body quantum states. In particular, the kinetic energy is constrained by charge density as $\braket{\hat{T}} \ge C_{ke}\int_{\vec{r}} n^{5/3}(\vec{r})$ \cite{kineticlieb}. Using the fact that $\braket{\hat{T}}$ is due to momentum fluctuations only, see Eq~\ref{eq:ke}, we obtain \begin{equation} \braket{T} = 3N\Delta p^2/2 \ge C_{ke}\int_{\vec{r}} n^{5/3}(\vec{r}) \end{equation} Instead of this bound, one could also obtain momentum fluctuations using (trial) Kohn-Sham wavefunctions, and improve bounds on momentum fluctuations iteratively. Also note that the constant $C_{ke}$ depends on the (anti)symmetry and spin-degeneracy of the system\cite{lieb1991inequalities}\footnote{Specifically, the $C_{ke}$ for a bosonic system varies by a factor $\propto N^{-2/3}$ from that of a fermionic system.}. With momentum fluctuations available, we can now use Equation~\ref{eq:uncertainb}, to constrain fluctuations of the electric field, and making that substitution, we obtain the total energy as a functional of $n(\vec{r})$ as follows \footnote{Could we have used constraints on $\braket{V_{ee}}$ \cite{lieb1981improved,chan1999optimized} first, and then applied Equation~\ref{eq:uncertain} to constrain momentum fluctuations? This approach is not as fruitful: bounds on $\braket{\hat{V}_{ee}}$ do not distinguish between fermionic and bosonic systems \cite{noteAnti2}; more importantly, bounds on $\braket{V_{ee}}$ do not contain local information necessary to constrain $\Delta E_{\parallel}^2(\vec{r})$.}
\begin{equation}
\mathscr{E}[n(\vec{r})] \sim C_{ke}\int n^{5/3}(\vec{r}) + \frac{\pi}{2 C_{ke}N} \frac{\int n^2 }{ \int n^{5/3}} +  \frac{1}{2} \int \frac{n(\vec{r}) n(\vec{r'})}{|\vec{r} - \vec{r'}|}+ \int V_{ext}(\vec{r}) n(\vec{r})
\label{eq:nonuniform}
\end{equation}

\textbf{\emph{Energy in the uniform density limit}} -- In the uniform density limit, $n(\vec{r}) = n$ and $V_{ext}(\vec{r}) = -\int_{\vec{r'}} \frac{n}{|\vec{r} - \vec{r'}|}$, a uniform positively charged background. The total energy then is: 
\begin{equation}
\mathscr{E}[n(\vec{r})] \sim C_{ke} n^{5/3} \mathcal{V} + \frac{\pi}{2 C_{ke}N} \frac{ n^2 \mathcal{V}}{ n^{5/3} \mathcal{V} }   + \frac{n^2}{2} \int \frac{1}{|\vec{r} - \vec{r'}|} d^3\vec{r} d^3\vec{r'} - {n^2} \int \frac{1}{|\vec{r} - \vec{r'}|} d^3\vec{r} d^3\vec{r'}  %- 4\pi N \Lambda 
\end{equation}
where $\mathcal{V}$ denotes volume of the system; let us use $n\mathcal{V} = N$ to substitute $\mathcal{V}$ in the first term on the \emph{right-hand-side}. Then, in terms of the Wigner-Seitz radius, $\frac{4\pi}{3}r_s^3 = \frac{1}{n}$, the energy is given by
\begin{equation}
\mathscr{E}^{uniform}[n] \sim NC_{ke} \left( \frac{3}{4\pi r_s^3}\right)^{2/3} +  \frac{\pi}{2NC_{ke}} \left(\frac{3}{4\pi r_s^3} \right)^{1/3}  + A(\mathcal{V})/2  \left( \frac{3}{4\pi r_s^3}\right)^2 + c_1(N)
\label{eq:bf_a}
\end{equation}
where we have added a constant term, $c_1(N)$ independent of $r_s$, but dependent on the total number of particles, $N$ (note the self-interaction constant that we ignored earlier). Rearranging in decreasing powers of $r_s$, we get:
\begin{equation}
\mathscr{E}^{uniform}[n(\vec{r})] \sim c_1(N)  + c_2(N)\frac{1}{r_s}  + c_3(N)\frac{1}{r_s^2} + c_4(\mathcal{V}) \frac{1}{r_s^6}
\label{eq:bf}
\end{equation}
\end{subequations}

\textbf{\emph{How ``universal'' is the energy expression in the uniform density limit?}} -- Equation~\ref{eq:bf} shows that bosonic and fermionic systems share the same $r_s$ dependencies despite (anti)symmetry or spin-degeneracy of $\Psi$. By linearity, we expect energy-differences between states that satisfy Equation~\ref{eq:bf} to also satisfy the same $r_s$ dependencies. Let us now fit Equation~\ref{eq:bf} against reference values of energies and energy-differences of different quantum states investigated using Quantum Monte Carlo\cite{ceperley1980ground}. 
\begin{figure}[htb!]
\centering \scriptsize
\begin{tabular}{cccc}
\hline
Coefficients for Eq.~\ref*{eq:bf}: & & & \\ 
\tiny \{-0.0041, -1.2336,  0.7985, -2.0181\} &  \tiny \{-0.0039, -1.2624,  2.5656, -0.1254 \} & \tiny \{ -0.0034,  -1.3089,   3.5297, 1.7527 \} & \tiny  \{ -0.0043, -1.2026,  0.8972 , 0 \} \\  \hline
Maximum Absolute Errors: & & & \\ 
1.43 mHa &  1.36 mHa & 1.26 mHa & 1.14 mHa \\ 
\hline \\
{\includegraphics[width=0.2\linewidth]{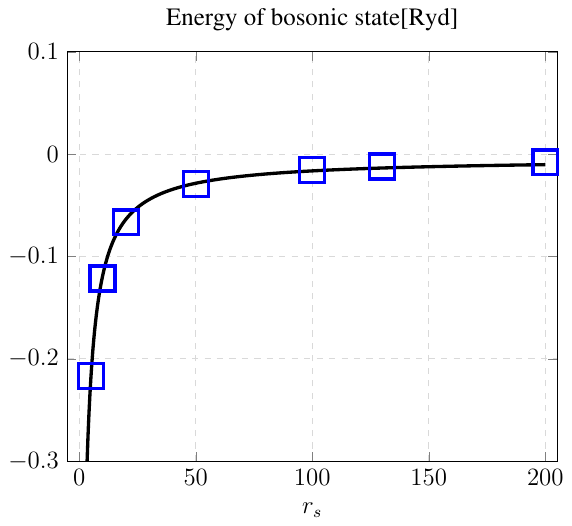}} & {\includegraphics[width=0.2\linewidth]{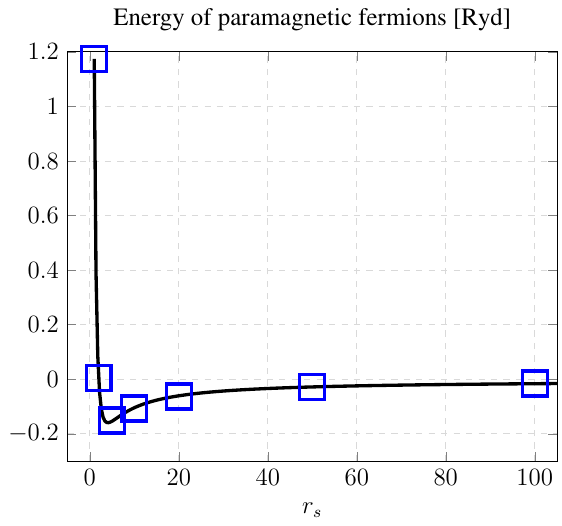}} & {\includegraphics[width=0.2\linewidth]{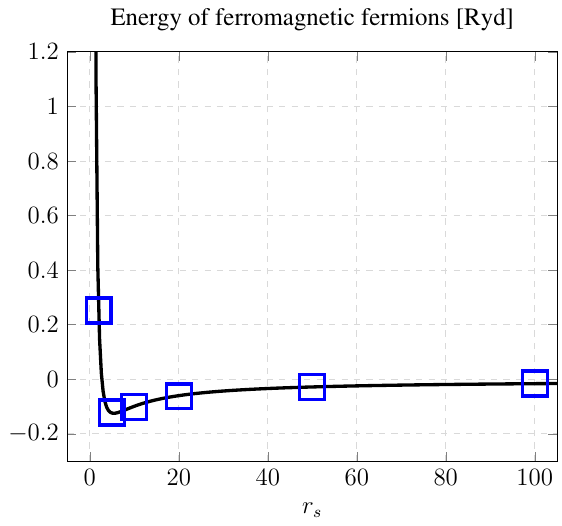}}  & {\includegraphics[width=0.2\linewidth]{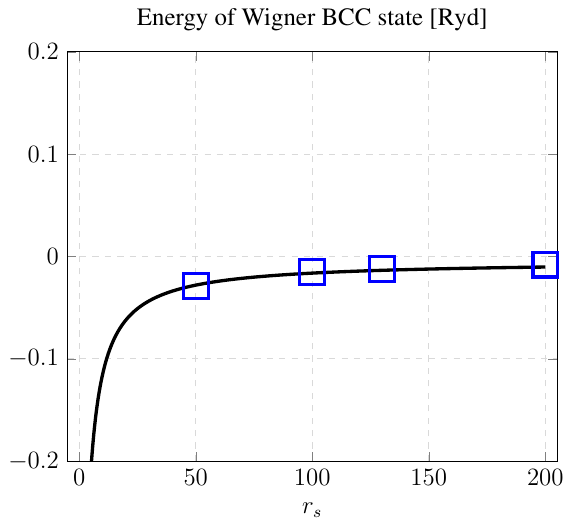}} \\ 
{\includegraphics[width=0.2\linewidth]{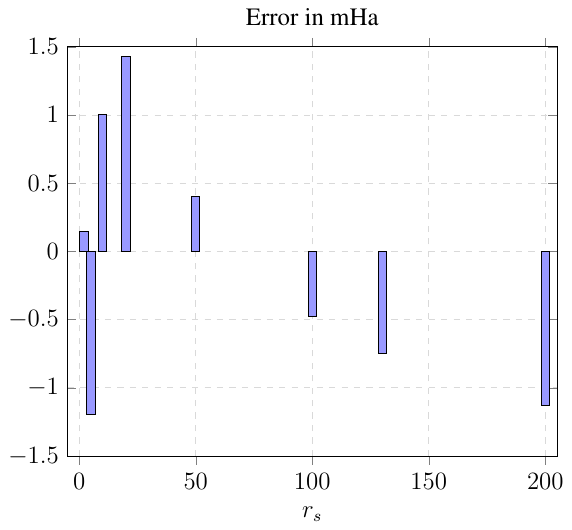}} & {\includegraphics[width=0.2\linewidth]{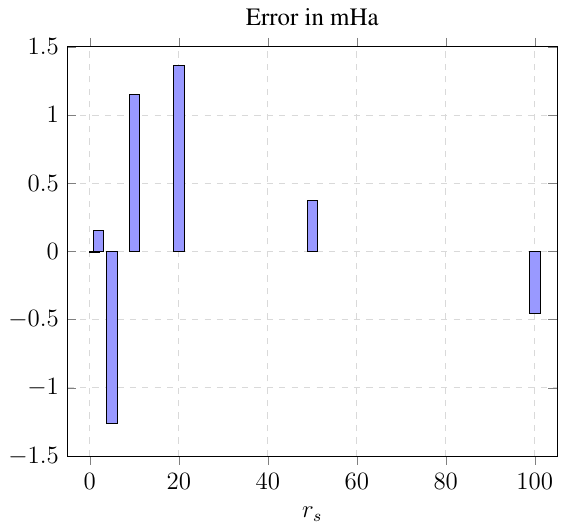}} & {\includegraphics[width=0.2\linewidth]{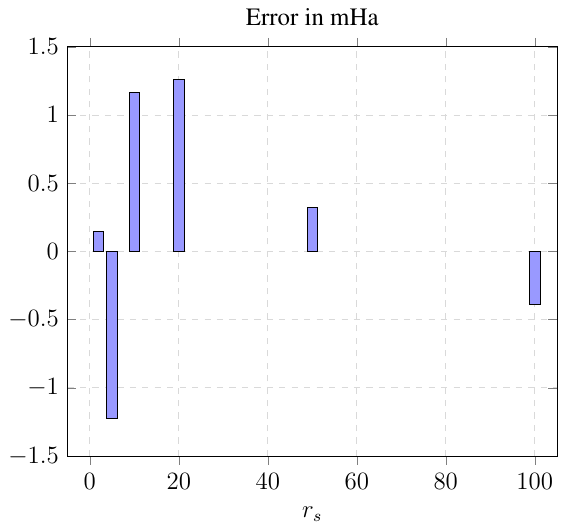}} & {\includegraphics[width=0.2\linewidth]{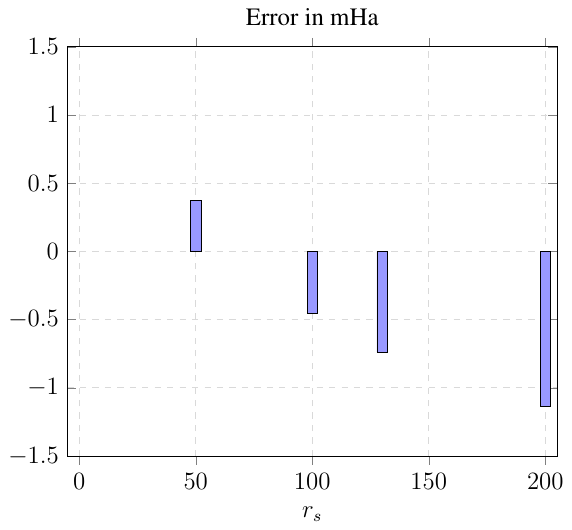}} \\
(a) & (b) & (c) & (d) \\
\end{tabular}
\caption{Consistency of Equation~\ref*{eq:bf} with Quantum Monte Carlo calculations of four types of many-body quantum wavefunctions in the uniform density limit \cite{ceperley1980ground}: (a) bosonic $\Psi$ (b) paramagnetic fermion (c) ferromagnetic fermion and (d) the Wigner BCC state. The first and second rows show coefficients obtained via semi-emprical fits of Equation~\ref*{eq:bf}, and corresponding maximum absolute deviation from QMC reference data in \cite{ceperley1980ground}. The third row shows energy as a function of $r_s$ where parameterized forms of Equation~\ref*{eq:bf} (solid lines) agree with QMC data (discrete points). Finally, deviation of energies from reference QMC data is plotted throughout the range of available data. Note that energies are in $\text{Ryd} = 0.5 \text{Ha}$ in keeping with \cite{ceperley1980ground}, and deviations are within $\pm 1.5$ mHa. These results support the general forms of $r_s$ dependence arising from uncertainty arguments in the previous section. \label{fig:energies}}
\end{figure}

Figure~\ref{fig:energies} shows the agreement of the functional form in Equation~\ref{eq:bf} to four types of quantum many-body states: bosonic fluid (BF), paramagnetic fermionic fluid (PMF), ferromagnetic fermion fluid (FMF), and Wigner crystalline state (BCC). Specifically, Figure~\ref{fig:energies} presents the coefficients that best describe each many-body quantum state, the maximum absolute deviation from reference data, comparison plots of energies (and deviations) of the functional form from QMC data for the complete range of $r_s$ values. Agreement of the functional form in Eq.~\ref{eq:bf} to reference values within $\pm 1.5\text{ mHa}$ suggests that we have captured the most essential physics implied by the kinetic and potential energy operators -- despite differences in the particle statistics, spin polarization, and crystallinity of different wavefunctions, they are described by the Hamiltonian (and resulting operator algebra) in Equation~\ref{eq:hamiltonian}.

\textbf{\emph{How well can we arrange many-body states by energy?}} -- We now consider whether Equation~\ref{eq:bf} is capable of representing transitions between quantum states. Energy-differences appear to generally agree with the $r_s$ dependencies, with notable exception around change in spin-polarization. 

\begin{figure}[htb!]
\begin{tabular}{ccc}
{\includegraphics[width=0.3\linewidth]{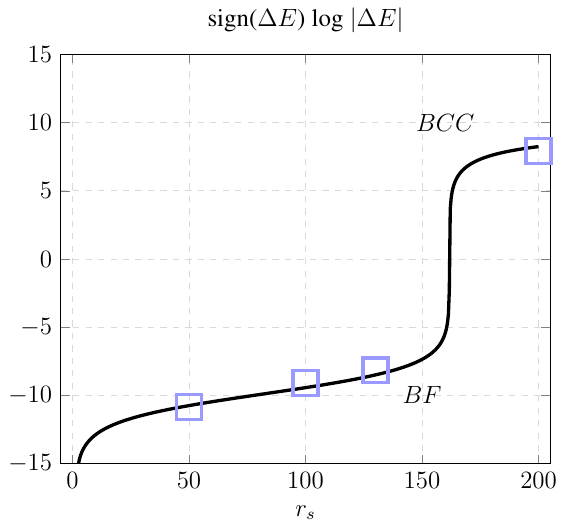}} & \includegraphics[width=0.3\linewidth]{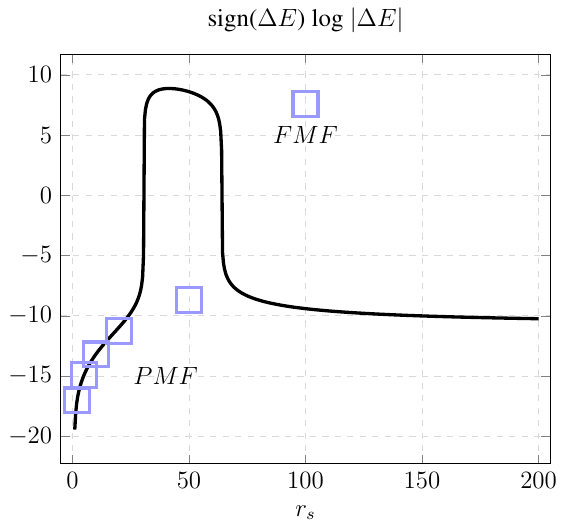} & \includegraphics[width=0.3\linewidth]{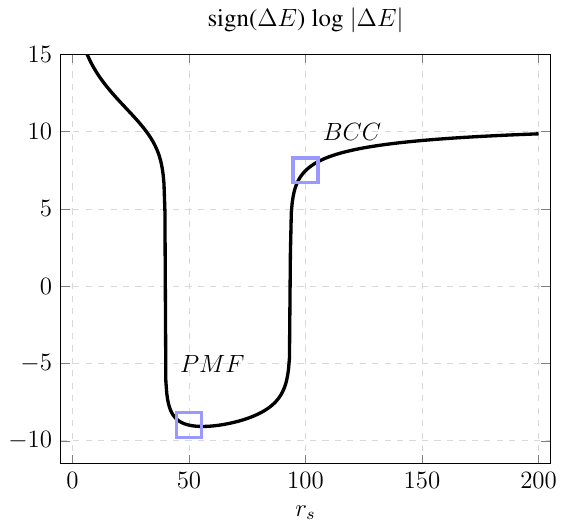} \\
\scriptsize (a) BF v. BCC Wigner state & \scriptsize  (b) PMF v. FMF & \scriptsize  (c) PMF v. BCC Wigner  \\
\end{tabular}
\caption{\label{fig:transitions} Energetic ordering of different quantum states in the jellium model, with discrete points showing QMC data\cite{ceperley1980ground}, and solid lines showing fits to Equation~\ref{eq:bf}, with coefficients from Figure~\ref{fig:energies}. Transitions occur when energy-differences change sign at $y=0$, with energetically favored states labeled before and after transition: (a) Transition between Bosonic fluid (BF) to BCC Wigner state is captured exactly at $r_s = 160$, (b) Transition between paramagnetic fermions (PMF) and ferromagnetic fermions (FMF) is not captured accurately, and (c) Transition between paramagnetic fermions (PMF) to BCC Wigner state is captured, despite the lack of reference datapoints \cite{ceperley1980ground}. As expected, the functional form does not capture spin-orbit coupling, absent in the Hamiltonian, while accurately representing the competition between kinetic and interaction energies that drive transitions around the Wigner state. Data from other QMC simulations are not combined, as each QMC simulation assumes extensivity, Jastrow factors, and sampling heuristics, with subtle effects that are beyond the scope of this work.}
\end{figure}

Figure~\ref{fig:transitions} shows energy differences between pairs of states. Specifically, in Figure~\ref{fig:transitions}a, the bosonic fluid (BF) to Wigner (BCC) state transition is captured exactly at $r_s=160$, in agreement with \cite{ceperley1980ground}, while in Figure~\ref{fig:transitions}b,  the functional form is unable to capture energy differences (and the transition) between paramagnetic and ferromagnetic states, despite high accuracy in energy prediction (\emph{cf.} Fig.~\ref{fig:energies}). This disagreement is promising (and expected) because while we have captured the competition between kinetic and interaction energies in the system, spin-orbit coupling, which crucial to determine the paramagnetic-to-ferromagnetic transition is not included. Finally, in Figure~\ref{fig:transitions}c, we observe that despite the lack of datapoints, the transition between Wigner (BCC) and paramagnetic fermion (PMF) states is correctly captured, as it is driven by competition between kinetic and interaction energies. More data from other sources are not used to complement the reference dataset in \cite{ceperley1980ground}, as each Quantum Monte Carlo simulation assumes size-extensivity, sampling heuristics, and trial wavefunctions that can lead to subtle effects, for instance see \cite{needs2020variational}. 

\textbf{Conclusion}
In this work, we have illuminated the pivotal role of the underlying electromagnetic field in the Hohenberg-Kohn theorem. It is shown that we can construct effective canonical variables for interacting many-body Hamiltonians, obtain uncertainty relations, and understand mechanistically, how nature implements the Hohenberg-Kohn theorem. Semi-empirical fits to QMC simulations in the uniform density limit agree with our conceptual proposal, but more work is necessary to develop universal density functionals based on the insights of this work, to conduct universality tests on benchmark non-uniform systems, and to handle periodicity. 

We can now understand the Hamiltonian in Equation~\ref{eq:hamiltonian} in terms of quantum fluctuations; one can similarly understand electron-phonon interactions in terms of quantum fluctuations of the external potential; understanding spin-orbit coupling and quantum magnetism, however, remains a challenge. More work along this direction would enable us to make general, material-agnostic, statements about the connection between antiferromagnetism and quantum phases such as strange-metals, high-temperature superconductors, and other non-fermi liquids\cite{nonfermiliquid}.%The theoretical failure of the quasiparticle or factorizability ansatz to describe strongly correlated systems, and the resulting transport properties 

\textbf{Acknowledgements} 
This material is based in part upon work supported by the Department of Energy, National Nuclear Security Administration, under Award Number DE-NA0002374, in collaboration with Prof. Harley T. Johnson at the University of Illinois.  Discussions with Johnson, David Ceperley, Andre Schleife and Lucas Wagner at University of Illinois are all gratefully acknowledged.  This work is also based in part upon research conducted at the Center for High Energy X-ray Sciences (CHEXS) which is supported by the National Science Foundation under award DMR-1829070. Discussions with Jacob Ruff at the Cornell High Energy Synchrotron Source (CHESS) are  gratefully acknowledged. 

%\newpage

%\appendix
% In our work, in addition to energy minimization, we seek a local principle that enforces this minimization, analogous to the global principle of action minimization of action of path integrals and corresponding local equations of motion. 
%\section{Time in many-particle quantum mechanics}

\bibliographystyle{unsrt}
\bibliography{paperNotes.bib}

\begin{thebibliography}{10}

\bibitem{hohenberg1964inhomogeneous}
Pierre Hohenberg and Walter Kohn.
\newblock Inhomogeneous electron gas.
\newblock {\em Physical Review}, 136(3B):B864, 1964.

\bibitem{chan1999optimized}
Garnet Kin-Lic Chan and Nicholas~C Handy.
\newblock Optimized {Lieb-Oxford} bound for the exchange-correlation energy.
\newblock {\em Physical Review A}, 59(4):3075, 1999.

\bibitem{lieb1979lower}
Elliott~H Lieb.
\newblock A lower bound for {Coulomb} energies.
\newblock {\em Physics Letters A}, 70(5-6):444--446, 1979.

\bibitem{lieb1981improved}
Elliott~H Lieb and Stephen Oxford.
\newblock Improved lower bound on the indirect coulomb energy.
\newblock {\em International Journal of Quantum Chemistry}, 19(3):427--439,
  1981.

\bibitem{liebkineticprl}
Elliott~H. Lieb and Walter~E. Thirring.
\newblock Bound for the kinetic energy of fermions which proves the stability
  of matter.
\newblock {\em Physical Review Letters}, 35:687--689, Sep 1975.

\bibitem{sham1966one}
LJ~Sham and Walter Kohn.
\newblock One-particle properties of an inhomogeneous interacting electron gas.
\newblock {\em Physical Review}, 145(2):561, 1966.

\bibitem{kohn1965self}
Walter Kohn and Lu~Jeu Sham.
\newblock Self-consistent equations including exchange and correlation effects.
\newblock {\em Physical Review}, 140(4A):A1133, 1965.

\bibitem{perdew2001jacob}
John~P Perdew and Karla Schmidt.
\newblock Jacob's ladder of density functional approximations for the
  exchange-correlation energy.
\newblock In {\em AIP Conference Proceedings}, volume 577, pages 1--20. AIP,
  2001.

\bibitem{mardirossian2015mapping}
Narbe Mardirossian and Martin Head-Gordon.
\newblock Mapping the genome of meta-generalized gradient approximation density
  functionals: The search for {B97M-V}.
\newblock {\em {The Journal of Chemical Physics}}, 142(7):074111, 2015.

\bibitem{tao2003climbing}
Jianmin Tao, John~P Perdew, Viktor~N Staroverov, and Gustavo~E Scuseria.
\newblock Climbing the density functional ladder: Nonempirical
  meta--generalized gradient approximation designed for molecules and solids.
\newblock {\em Physical Review Letters}, 91(14):146401, 2003.

\bibitem{kohn1983v}
Walter Kohn.
\newblock V-representability and density functional theory.
\newblock {\em Physical Review Letters}, 51(17):1596, 1983.

\bibitem{yang2004potential}
Weitao Yang, Paul~W Ayers, and Qin Wu.
\newblock Potential functionals: dual to density functionals and solution to
  the v-representability problem.
\newblock {\em Physical Review Letters}, 92(14):146404, 2004.

\bibitem{lieb1983eh}
Elliott~H Lieb.
\newblock {Density Functionals for CouIomb Systems}.
\newblock {\em International Journal of Quantum Chemistry}, 24:243, 1983.

\bibitem{nonfermiliquid}
Sung-Sik Lee.
\newblock Recent developments in non-fermi liquid theory.
\newblock {\em Annual Review of Condensed Matter Physics}, 9(1):227--244, 2018.

\bibitem{straying}
Michael~G. Medvedev, Ivan~S. Bushmarinov, Jianwei Sun, John~P. Perdew, and
  Konstantin~A. Lyssenko.
\newblock Density functional theory is straying from the path toward the exact
  functional.
\newblock {\em Science}, 355(6320):49--52, 2017.

\bibitem{dilemma_on}
Sharon Hammes-Schiffer.
\newblock A conundrum for density functional theory.
\newblock {\em Science}, 355(6320):28--29, 2017.

\bibitem{comment_on}
Kasper~P. Kepp.
\newblock Comment on ``{Density functional theory is straying from the path
  toward the exact functional}''.
\newblock {\em Science}, 356(6337):496--496, 2017.

\bibitem{ceperley1978ground}
David~M Ceperley.
\newblock Ground state of the fermion one-component plasma: A {Monte Carlo}
  study in two and three dimensions.
\newblock {\em Physical Review B}, 18(7):3126, 1978.

\bibitem{ceperley1980ground}
David~M Ceperley and BJ~Alder.
\newblock Ground state of the electron gas by a stochastic method.
\newblock {\em Physical Review Letters}, 45(7):566, 1980.

\bibitem{mitas2006structure}
Lubos Mitas.
\newblock Structure of fermion nodes and nodal cells.
\newblock {\em Physical Review Letters}, 96(24):240402, 2006.

\bibitem{ceperley1991fermion}
David~M Ceperley.
\newblock Fermion nodes.
\newblock {\em Journal of Statistical Physics}, 63(5-6):1237--1267, 1991.

\bibitem{carlson2011auxiliary}
J~Carlson, Stefano Gandolfi, Kevin~E Schmidt, and Shiwei Zhang.
\newblock Auxiliary-field quantum monte carlo method for strongly paired
  fermions.
\newblock {\em Physical Review A}, 84(6):061602, 2011.

\bibitem{sunko2016natural}
DK~Sunko.
\newblock Natural generalization of the ground-state slater determinant to more
  than one dimension.
\newblock {\em Physical Review A}, 93(6):062109, 2016.

\bibitem{note2b}
Use $\int \frac{1}{|\vec{k}|^2} d^3\vec{k} = \lim_{m\rightarrow 0}
  \int_{k}\frac{1}{k^2 + m^2} d^3\vec{k}$. In D dimensions, the integral is
  generalized to \begin{align} I(D) &= \int \frac{d^Dk}{(2\pi)^D}
  \frac{1}{\vec{k}^2 + m^2} \nonumber \\ & = \frac{2\pi}{(2\pi)^D}
  \prod_{k=1}^{D-2} \int_0^{\pi} \sin^k\theta_k d\theta_k \int_0^{\infty} dk
  k^{D-1}\frac{1}{p^2 + m^2} \nonumber \\ & = \frac{S_D}{(2\pi)^D}
  \int_0^{\infty} dk k^{D-1} \frac{1}{k^2 + m^2} \nonumber \end{align} where
  $S_D = \frac{2\pi^{D/2}}{\Gamma(D/2)}$ is the surface of a unit sphere in D
  dimensions. Substituting $k^2 = ym^2$, the resulting integral can be written
  as : $B(\alpha, \gamma) = \frac{\Gamma(\alpha) \Gamma(\gamma)}{\Gamma(\alpha
  + \gamma)} = \int_0^{\infty}dy y^{\alpha-1}(1+y)^{-\alpha - \gamma}$ We find
  that: \[ \lim_{m\rightarrow 0} I(D) = \frac{ (m^2)^{D/2-1}}{(4\pi)^{D/2}}
  \Gamma(1 - D/2) = 0 \].

\bibitem{lieb1988Denmark}
Elliott~H Lieb.
\newblock Kinetic energy bounds and their application to the stability of
  matter.
\newblock In {\em Inequalities, Selecta of Elliot H Lieb, Eds. Loss and
  Ruskai}, pages 317--328. Springer, 2002.

\bibitem{robertson1929uncertainty}
Howard~Percy Robertson.
\newblock The uncertainty principle.
\newblock {\em Physical Review}, 34(1):163, 1929.

\bibitem{ryder1996quantum}
Lewis~H Ryder.
\newblock {\em Quantum field theory}.
\newblock Cambridge university press, 1996.

\bibitem{kineticlieb}
Elliott~H. Lieb and Walter~E. Thirring.
\newblock Bound for the kinetic energy of fermions which proves the stability
  of matter.
\newblock {\em Phys. Rev. Lett.}, 35:687--689, Sep 1975.

\bibitem{lieb1991inequalities}
Elliott~H Lieb and Walter~E Thirring.
\newblock Inequalities for the moments of the eigenvalues of the {Schrodinger
  Hamiltonian and their relation to Sobolev inequalities}.
\newblock In {\em The Stability of Matter: From Atoms to Stars}, pages
  135--169. Springer, 1991.

\bibitem{Note1}
Specifically, the $C_{ke}$ for a bosonic system varies by a factor $\propto
  N^{-2/3}$ from that of a fermionic system.

\bibitem{Note2}
Could we have used constraints on $\mathinner {\langle {V_{ee}}\rangle }$ \cite
  {lieb1981improved,chan1999optimized} first, and then applied Equation~\ref
  {eq:uncertain} to constrain momentum fluctuations? This approach is not as
  fruitful: bounds on $\mathinner {\langle {\protect \hat {V}_{ee}}\rangle }$
  do not distinguish between fermionic and bosonic systems \cite {noteAnti2};
  more importantly, bounds on $\mathinner {\langle {V_{ee}}\rangle }$ do not
  contain local information necessary to constrain $\Delta E_{\parallel
  }^2(\protect \mathbf {r})$.

\bibitem{needs2020variational}
RJ~Needs, MD~Towler, ND~Drummond, Pablo Lopez~Rios, and JR~Trail.
\newblock Variational and diffusion quantum monte carlo calculations with the
  casino code.
\newblock {\em The Journal of Chemical Physics}, 152(15):154106, 2020.

\bibitem{noteAnti2}
Here we elaborate on previous Lieb-Thirring and Lieb-Oxford type bounds on
  kinetic and interaction energies respectively, where the effect of
  antisymmetry is somewhat tangentially discussed. Suppose we have a many-body
  solution for interacting systems with distinguishable particle labels,
  $\Psi_C(x_1, \ldots, x_N)$, that results in a charge density, $n(\vec{r})$.
  One can construct a symmetric (bosonic) many-body wavefunction, $\Psi_S$, by
  summing over permutations of the coordinate labels, \[ \Psi_S(x_1, \ldots,
  x_N) =\frac{1}{\sqrt{N!}} \sum_{\mathscr{P}(x_1, \ldots, x_N)}
  \Psi(\mathscr{P}) \] To obtain a corresponding antisymmetric (fermionic)
  many-body wavefunction resulting in the same charge density, $n(\vec{r})$,
  $\Psi_A$, we multiply the symmetric wavefunction by $\Theta(x_1, \ldots,
  x_N)$, where $\Theta = \pm 1$ represents the fermion nodal structure and
  changes sign under exchange of coordinate labels, $\Psi_A = \Theta(x_1,
  \ldots, x_N) \Psi_S(x_1, \ldots, x_N)$. In the following, subscripts $S$
  denote observables of symmetric wavefunctions while $A$ denote observables
  corresponding to antisymmetric wavefunctions. We find that expectation values
  of operators like $\delta(\vec{r} - x_\alpha)$, and $\frac{1}{|x_\alpha -
  x_\beta|}$ remain unchanged because for antisymmetric wavefunctions, the
  integrals simply include $\Theta^2 = 1$: \[ n_{A}(\vec{r}) = \int_{x_1,
  \ldots, x_N} |\Psi_A|^2 \delta(\vec{r} - x_\alpha) = \int_{x_1, \ldots, x_N}
  |\Psi_S|^2 \delta(\vec{r} - x_{\alpha}) \Theta^2 = n_{S}(\vec{r})\] \[
  V_{ee,A}(\vec{r}) = \int_{x_1, \ldots, x_N} |\Psi_A|^2 \frac{1}{|x_\alpha -
  x_\beta|} = \int_{x_1, \ldots, x_N} |\Psi_S|^2 \frac{1}{|x_\alpha - x_\beta|}
  \Theta^2 = V_{ee,S}(\vec{r})\] On the other hand, the momentum operator
  requires the derivative of $\Psi$, and the presence of nodal structure
  becomes relevant. Specifically, for $\Psi_S \rightarrow \Psi_A$ related by
  the fermion nodal structure, the observed momenta vary because:
  \[\braket{p}_{\alpha, A} = \frac{\hbar}{i} \int_{x_1, \ldots, x_N}
  \Psi_A^\dagger(x_1, \ldots, x_N) \nabla_\alpha \Psi_A (x_1, \ldots, x_N)\]
  Operating the derivative on $\nabla_\alpha \Psi_A = (\nabla_\alpha \Psi_S)
  \Theta + \Psi_S \nabla_\alpha \Theta $, resulting in \[ \braket{p}_{\alpha,
  A} = \braket{p}_{\alpha, S} + \frac{\hbar}{i}\int_{x_1, \ldots, x_N}
  |\Psi_S|^2 \Theta(x_1, \ldots, x_N) \nabla_\alpha \Theta(x_1, \ldots, x_N) \]
  In case of the local electric field operator, the gradient is with respect to
  $\vec{r}$ acting on $\frac{1}{|\vec{r} - \vec{x}_1|}\Psi(x_1, \ldots, x_N)$,
  and derivatives on the nodal-surface do not appear.

\end{thebibliography}

\end{document}